\documentclass{ws-ijmpd}

\begin{document}

\markboth{A. Accioly, J. Helay\"{e}l-Neto and E. Scatena}
{Photon Mass and  Very Long Baseline Interferometry}

%
\catchline{}{}{}{}{}
%

\title{PHOTON MASS AND  VERY LONG BASELINE INTERFEROMETRY\footnote{This article is an extended version of the essay `COMBINING GENERAL RELATIVITY, MASSIVE QED AND  VERY LONG BASELINE INTERFEROMETRY TO GRAVITATIONALLY CONSTRAIN THE PHOTON MASS' (A. Accioly, J. Helay\"{e}l-Neto and E. Scatena, {\it Phys. Lett. A} {\bf374} (2010) 3806) which was awarded an ``honorable mention" in the 2010 Essay Competition of the Gravity Research Foundation.}  }

\author{\footnotesize ANTONIO ACCIOLY$^{\dagger,\; \star,\; \amalg}$, \; JOS\'{E} HELAY\"{E}L-NETO$^{\dagger,\; \diamond}$ and ESLLEY SCATENA$^{\star, \; \ddagger}$}

\address{$^{\dagger}$Laborat\'{o}rio de F\'{\i}sica Experimental (LAFEX), \\Centro Brasileiro de Pesquisas  F\'{\i}sicas (CBPF),\\ Rua Dr. Xavier Sigaud 150, Urca, 22290-180, \\Rio de Janeiro, RJ, Brazil\\ 
$^{\star}$Instituto de F\'{\i}sica Te\'{o}rica (IFT), \\ S\~ao Paulo State University (UNESP), \\R. Dr. Bento Teobaldo Ferraz 271, Bl. II - Barra Funda, \\01140-070, S\~ao Paulo, SP, Brazil\\    $^{\amalg}$accioly@cbpf.br\\$^{\diamond}$helayel@cbpf.br\\$^{\ddagger}$scatena@ift.unesp.br }

\maketitle

\begin{history}
\received{Day Month Year}
\revised{Day Month Year}
\comby{Managing Editor}
\end{history}

\begin{abstract}
A relation between the photon mass, its frequency, $\nu$, and the deflection parameter, $\gamma$, determined by experimentalists (which characterizes the contribution of space curvature to gravitational deflection) is found. This amazing result allows us to conclude that the knowledge of the parameters $\nu$ and $\gamma$ is all we need to set up gravitational bounds on the photon mass. By considering as inputs the most recent measurements of the solar gravitational deflection of radio waves obtained via the Very Long Baseline Interferometry, upper bounds on the photon mass are estimated.    
\end{abstract}

\section{Introduction}	

As a consequence of the substantial improvement in the interferometry techniques that took place over the last few decades, a new light was shed on the physical phenomena in which gravitational and quantum effects are interwoven. Consider, for instance, the  quantum mechanical phase shift of neutrons caused by their interaction with Earth's gravitational field which was observed in the Colella-Overhauser-Werner\cite{1} and  Bonse-Wrobleski\cite{2} experiments performed with neutron interferometry. These seminal and pioneering scientific tests  have helped dramatically to convince physicists that there exist  some experiments whose outcome necessarily depends upon both the gravitational constant and the Planck's constant. The fact that in the great majority of phenomena of interest in terrestrial physics, gravity and quantum mechanics do not {\it simultaneously} play an important role, is certainly responsible for the wrong idea that  gravity and quantum mechanics cannot be closely intertwined in some special circumstances. In the last four decades the COW experiments have become more sophisticated. The latest  neutron interferometry experiments\cite{3} report a statistically significant discrepancy between the theoretically predicted and experimentally measured values of the neutron phase shift due to gravity. Supposing the equality of  inertial and  gravitational masses for the neutron the experimenters found the  neutron phase factor to be $ 1\% $ lower than  predicted,\cite{3} which clearly signals a possible violation of the classical equivalence principle. At first sight, it seems  that  these latest neutron interferometry experiments are in conflict   with the more precise tests of the classical equivalence principle conducted via atomic interferometry, and with those based on torsion pendulum. Adunas, Rodriguez-Milla and Ahluwalia,\cite{4} showed, however, that each of the aforementioned  experiments probes a different aspect of gravity; and that current experiment techniques, when coupled to solar-neutrino data, may be able to explore quantum mechanically induced violations of the classical equivalence principle. They also predicted a quantum violation of the classical equivalence principle for the  next generation of atomic interferometry experiments. Actually, from an operational point of view one cannot claim, even in principle, that there exists, for certain quantum systems, an exact equality of gravitational  and inertial masses.\cite{4,5} Therefore, we come to the conclusion that quantum mechanics and the classical equivalence principle cannot coexist peacefully.\cite{6,7}                     

Recently,  an interesting experiment, not directly related to interferometry techniques, but that attests to the fact that neutrons can reside in quantum stationary states formed in the gravitational field of the Earth,  was carried out. In this experiment, the lowest quantum state of neutrons in the Earth's gravitational field was identified in the measurement of neutron transmission between a horizontal mirror on the bottom and an absorber/scatter on top.\cite{8,9} This result motivate  Ernest to do a careful and through research on   gravitational eigenstates in weak gravity.\cite{10,11} Actually, despite the almost universal study of quantum theory applied to atomic and molecular states, very little work has been done to investigate the properties of the hypothetical stationary states that should exist in similar types of gravitational central potentials wells, particularly those with large quantum numbers.

The preceding experiments show clearly  that physical phenomena in which gravity (specifically, Earth's gravitational field)  and quantum effects are merged together, are no more beyond our reach. We remark that in all these investigations the  gravitational field of the Earth is described by Newton's gravity.

 On the other hand,  since its original publication in 1915, Einstein's general theory of  relativity continues to be an active area of both theoretical and experimental research. Presently, the theory successfully accounts for all data gathered to date.\cite{12} In addition, among its so-called classical tests there is one, namely, light bending, that has been confirmed with an accuracy that increases as time goes by. In reality, it is expected that a series of improved designed experiments with the Very Long Baseline Interferometry (VLBI) will increase the present accuracy of the deflection parameter $\gamma$ by at least a factor of 4.\cite{13} By the way, the current value for $\gamma$ is 0.9998$\pm$0.0003 (68$\%$ confidence level),\cite{13} in agreement with general relativity. Besides, it is also a generally acknowledged fact that the gravitational deflection of light by the sun can be measured more accurately at radio wavelengths with interferometry  techniques than at visible wavelength with available optical techniques.$^{14-17}$ Indeed, at present the VLBI is the most accurate technique we have at our disposal for measuring radio-wave gravitational deflection.\cite{13} Now, taking into account that the search for upper bounds on the photon mass\footnote{In all these researches, the photon is described by massive QED, which is nothing but the most straightforward extension of standard QED. Its Lagrangian can be written as \begin{equation} \mathcal{L}=-\frac{1}{4}F^{2}_{\mu\nu} + \frac{1}{2}m^2A^{2}_{\mu}- J^{\mu}A_{\mu},\end{equation} where $F_{\mu\nu}$($=\partial_{\mu}A_{\nu}-\partial_{\nu}A_{\mu}$) is the field strength, $J^{\mu}$ is the conserved (electric) current and $m$ is the photon mass. Here, indices are raised and lowered with $\eta^{\mu\nu}$ and $\eta_{\mu\nu}$, respectively. In reality, massive QED is theoretically simpler than the standard theory,\cite{18} besides being renormalizable.\cite{19}} has increased over the past several decades,$^{20-22}$ it would be interesting to estimate gravitational bounds for the photon mass by considering the most recent measurements of the solar  gravitational deflection of radio waves obtained by means of the VLBI. This is precisely  the goal of this work.

The article is organized as follows. In Sec. 2 we show that there exists a constraint on the photon mass, its frequency and the deflection parameter determined by experimenters (which characterizes the contribution of space curvature to gravitational deflection). From this amazing result, upper limits on the photon mass are found in Sec. 3, by considering as inputs the most recent measurements of the solar gravitational deflection of radio waves obtained via the VLBI.\cite{13} To conclude, we discuss in Sec. 4, whether or not the bounds we have  estimated can be improved.

In our convention $\hbar = c= 1$, and the signature is (+ - - -). 

\section{Combining General Relativity, Massive QED and  Very Long Baseline Interferometry to Gravitationally Constrain the Photon Mass   }

To start off, we recall that the Lagrangian for the gravitational minimally coupled massive photon field is

\begin{eqnarray}
{\cal{L}} = \sqrt{-g} \left[-\frac{1}{4}g^{\mu \alpha} g^{\nu \beta}F_{\mu \nu}F_{\alpha \beta} + \frac{m^2}{2} g^{\mu \nu} A_\mu A_\nu  \right].
\end{eqnarray}

On the other hand, for  small fluctuations  around the Minkowski metric $\eta$, the full metric can be written as  

\begin{figure}[t]
\centerline{\psfig{file=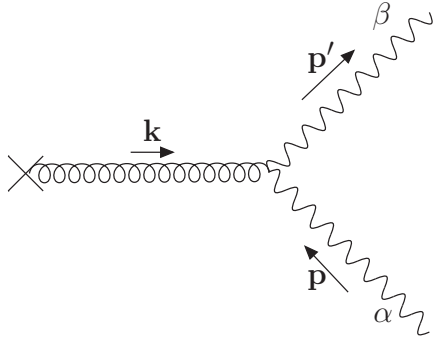,width=5.9cm}}
\caption{Feynman graph for the interaction between a massive photon  and an external gravitational field.\label{fig1}}
\end{figure}

\begin{eqnarray}
g_{\mu \nu}= \eta_{\mu \nu} + \kappa h_{\mu \nu},
\end{eqnarray}
 \noindent where $\kappa^2 = 32 \pi G$. Here $G$ is Newton's constant. 

Supposing then that the massive photon is scattered by an external weak gravitational field (See Fig. 1), we promptly obtain from Eqs. (2) and (3) that the  Lagrangian for the interaction under discussion is 

\begin{eqnarray}
{\cal{L}}_\mathrm{int} = \frac{\kappa}{2}h_\mathrm{ext}^{\mu \nu} \left[\eta^{\alpha \beta}F_{\alpha \mu}F_{\beta \nu} - \frac{1}{4}F_{\alpha \beta}^2 \eta_{\mu \nu} +  \frac{m^2}{2} A_\alpha^2 \eta_{\mu \nu} -m^2  A_\mu A_\nu  \right].
 \end{eqnarray} 

\noindent Note that now the indices are raised (lowered) with $ \eta^{\mu \nu} (\eta_{\mu \nu})$.

Accordingly, the vertex function for the alluded process is, in momentum space, given by

\begin{eqnarray}
\nonumber V_{\alpha\beta}(p,p')&=&\frac{\kappa}{2}h^{\mu\nu}_{\mathrm{ext}}(\mathbf{k})\Big[(m^2-p\cdot p')(\eta_{\mu\nu}\eta_{\alpha\beta}-2\eta_{\mu\alpha}\eta_{\nu\beta})+p'_{\alpha}p_{\beta}\eta_{\mu\nu}\\ &&+      \;2(-p'_{\alpha}p_{\nu}\eta_{\mu\beta}-p'_{\mu}p_{\beta}\eta_{\nu\alpha}+p'_{\mu}p_{\nu}\eta_{\alpha\beta})\Big]. \label{vertice1}
\end{eqnarray}

Now, assuming that the external weak gravitational field is generated by a point mass $M$ at $\bf {r}$=$\bf{0}$, we immediately find the expression for the external gravitational field  by solving  Einstein's linearized field equations   in  the de Donder gauge. The resulting expression is

\begin{equation}
h^{\mu\nu}_{\mathrm{ext}}(\mathbf{r})=\frac{\kappa M}{16\pi r}(\eta^{\mu\nu}-2\eta^{\mu0}\eta^{\nu0}).
\end{equation}

Consequently, the momentum space gravitational field, $h_\mathrm{ext}^{\mu \nu}({\bf k})$, is given by

\begin{eqnarray}
\nonumber h^{\mu\nu}_{\mathrm{ext}}(\mathbf{k})&=&\int{d^{3}\mathbf{r}e^{-i\mathbf{k}\cdot\mathbf{r}}h^{\mu\nu}_{\mathrm{ext}}(\mathbf{r})} \\&=&\frac{\kappa M}{2\mathbf{k}^2}\Big(\frac{\eta^{\mu\nu}}{2}-\eta^{\mu0}\eta^{\nu0}\Big)\label{hexterno}.
\end{eqnarray}

We are now ready to compute the unpolarized differential cross section for the process displayed in Fig. 1. To do that we recall that the expression for the mentioned cross section is

\begin{equation}
\frac{d\sigma}{d\Omega}=\frac{1}{(4\pi)^{2}}\frac{1}{3}\sum^{3}_{r=1}\sum^{3}_{r'=1}\mathcal{M}^{2}_{rr'},\label{schoque}
\end{equation}

\noindent with

\begin{equation}
{\cal{M}}_{rr'}=\epsilon_r^\alpha ({\bf p})\epsilon_{r'}^\beta ({\bf p'})V_{\alpha \beta}(p, p'),
\end{equation}

\noindent where $\epsilon_r^\alpha ({\bf p})$ and $\epsilon_{r'}^\beta ({\bf p'})$ are the polarizations vectors for the ingoing and outgoing vectorial bosons, respectively. These vectors, in turn,  satisfy the  relation

\begin{equation}
\sum_{r=1}^{3}\epsilon_{r}^{\mu}(\mathbf{p})\epsilon^{\nu}_{r}(\mathbf{p})=-\eta^{\mu\nu}+\frac{p^{\mu}p^{\nu}}{m^2}.\label{completeza}
\end{equation}

After algebraic manipulations, we find that the expression for the unpolarized  differential cross section we are searching for reads

\begin{equation} \frac{d\sigma}{d\Omega}=\frac{1}{6}\frac{M^2G^2}{\Big(\sin^{2}{\frac{\theta}{2}}\Big)^{2}}\Bigg[3\mathbf{p}^{4}+\frac{3}{2}m^{4}+2\mathbf{p}^{2}m^2+2\mathbf{p}^{2}(\mathbf{p}^{2}+2m^2)\cos{\theta}+\mathbf{p}^{4}\cos^{2}{\theta}\Bigg],
\end{equation}

\noindent where $\theta$ is the scattering angle. For small angles, it  reduces to

\begin{equation}
\frac{d\sigma}{d\Omega}=\frac{16M^2G^2}{\theta^4}\left[\frac{1-\frac{m^2}{2E^{2}}}{1-\frac{m^2}{E^{2}}}\right]^{2},
\end{equation}

\noindent where $E$ is the energy of the ingoing photon.

The above differential cross section can be related to a classical trajectory with  impact parameter $b$ via the relation 
\begin{eqnarray}
bdb = - \frac{d\sigma}{d\Omega} \theta d\theta.
\end{eqnarray} 

From Eqs. (12) and (13), we arrive at the conclusion that

\begin{eqnarray}
\theta = \frac{4MG}{b}\left(\frac{1 - \frac{m^2}{2E^2}}{1 - \frac{m^2}{E^2}}\right),
\end{eqnarray}

\noindent which, in the ultrarelativistic limit, i.e., $E \gg m$,  leads to

\begin{eqnarray}
\theta &=& \theta_{\mathrm E}\left(1 + \frac{m^2}{2E^2}\right)  \\ &=&  \theta_{\mathrm E}\left( 1 + \frac{m^2}{8 \pi^2 \nu^2}\right),
\end{eqnarray}

\noindent where  $\nu$ is  the frequency of the  massive photon and $ \theta_{\mathrm E} \equiv \frac{4MG}{b}$.

Before going on, some comments are in order. 

\begin{itemlist}

\item {Recently, it was shown that the unpolarized differential cross sections for the scattering of different quantum particles are spin dependent, which is in disagreement with the classical equivalence principle\cite{7} (See Table 1). This result raises an important question: Why the gravitational field perceives 
the spin? Because there is the presence of a nonzero momentum transfer ({\bf k}) in the scattering, responsible for probing the internal structure (spin) of the particle. Nevertheless, if we choose any two expressions from those listed in Table 1, we get that the difference between them is always extremely small for typical deflection angles. To show this for the massless particles, for instance, we study the behavior of 
\begin{eqnarray}
\frac{\Delta(\frac{d\sigma}{d\Omega})}{(\frac{d\sigma}{d\Omega})_{s=0}} \equiv \frac{(\frac{d\sigma}{d\Omega})_s - (\frac{d\sigma}{d\Omega})_{s=0}}{(\frac{d\sigma}{d\Omega})_{s=0}},
\end{eqnarray}

\noindent as a function of the scattering angle $\theta$ (See Fig. 2). It is trivial to show that for small angles the preceding expression reduces to

\begin{eqnarray}
\frac{\Delta(\frac{d\sigma}{d\Omega})}{(\frac{d\sigma}{d\Omega})_{s=0}} \approx  - \frac{s \theta^2}{2}.
\end{eqnarray}

\noindent For a typical deflection angle, say $\theta \sim 10^{-6}$, we found $\frac{\Delta(\frac{d\sigma}{d\Omega})}{(\frac{d\sigma}{d\Omega})_{s=0}} \sim 10^{-12}$.  The detection of so small an effect is, of course,  beyond todays technology. Consequently,  for  these tiny deflection angles, the cross sections will be unaffected by the spin of the particle.

\begin{table}[h]
\tbl{Unpolarized differential cross-sections for the scattering of different quantum particles by an external weak gravitational field generated by a static point particle of mass $M$. Here $m$ is the particle mass, $s$ the spin, $\theta$ the scattering angle, and $\lambda\equiv\frac{m^{2}}{\mathbf{p}^{2}}=\frac{1-\mathbf{v}^{2}}{\mathbf{v}^{2}}$, with $\mathbf{v}$ and $\mathbf{p}$ being the velocity and three-momentum, in this order, of the incident particle.}
{\begin{tabular}{ccc}
\hline
\hline
$m$&$s$&$\frac{d\sigma}{d\Omega}$\\
\hline
\\
0 & 0 & $\Big(\frac{GM}{\sin^{2}{\frac{\theta}{2}}}\Big)^{2}$\\
\\
$\neq 0$ & 0 & $\Big(\frac{GM}{\sin^{2}{\frac{\theta}{2}}}\Big)^{2}\Big(1+\frac{\lambda}{2}\Big)^{2}$\\
\\
0 & $\frac{1}{2}$ & $\Big(\frac{GM}{\sin^{2}{\frac{\theta}{2}}}\Big)^{2}\cos^{2}{\frac{\theta}{2}}$\\
\\
$\neq 0$ & $\frac{1}{2}$ & $\Big(\frac{GM}{\sin^{2}{\frac{\theta}{2}}}\Big)^{2}\Big[\cos^{2}{\frac{\theta}{2}}+\frac{\lambda}{4}\Big(1+\lambda+3\cos^{2}{\frac{\theta}{2}}\Big)\Big]$\\
\\
0 & 1 & $\Big(\frac{GM}{\sin^{2}{\frac{\theta}{2}}}\Big)^{2}\cos^{4}{\frac{\theta}{2}}$\\
\\
$\neq 0$ & 1 & $\Big(\frac{GM}{\sin^{2}{\frac{\theta}{2}}}\Big)^{2}\Big[\frac{1}{3}+\frac{2}{3}\cos^{4}{\frac{\theta}{2}}-\frac{\lambda}{3}\Big(1-\frac{3\lambda}{4}-4\cos^{2}{\frac{\theta}{2}}\Big)\Big]$\\
\\
0 & 2 & $\Big(\frac{GM}{\sin^{2}{\frac{\theta}{2}}}\Big)^{2}\Big(\sin^{8}{\frac{\theta}{2}}+\cos^{8}{\frac{\theta}{2}}\Big)$\\ 
\hline
\hline
\end{tabular}}
\end{table} }

\item{In order to recover Einstein's geometrical results from Table 1, we must have ${\bf k } \rightarrow {\bf 0}$; in other words, in the nontrivial limit of small momentum transfer, which corresponds to a nontrivial small angle limit since $\left| {\bf k}\right| = 2\left| {\bf p}\right| \sin\frac{\theta}{2}$, the massive (massless) particles behave in the same way, regardless the spin. In fact, when the spin is `switched off', i.e, for small angles, we obtain from Table 1 that for $m=0$,

\begin{eqnarray}
 \frac{d\sigma}{d\Omega} \approx \frac{16 G^2 M^2}{\theta^4},
 \end{eqnarray}
 
 \noindent while for $m \neq 0$,

\begin{eqnarray}
 \frac{d\sigma}{d\Omega} \approx \frac{16 G^2 M^2}{\theta^4}\left(1 + \frac{\lambda}{2}\right)^2.
 \end{eqnarray}

Using Eq. (12) we conclude that for $m=0$,

\begin{eqnarray}
 \theta \approx \frac{4 G M}{b},
 \end{eqnarray}
 
 \noindent while for $m \neq 0$,

\begin{eqnarray}
 \theta \approx \frac{4 G M}{b}\left(1 + \frac{\lambda}{2}\right).
 \end{eqnarray}

\noindent The former equation gives the gravitational deflection angle for a massless particle --- a result foreseen by Einstein a long time ago; whereas the latter coincides with the prediction of   general relativity for the deflection of  a massive classical test particle by an  external weak gravitational field.\cite{23} It is worth noticing that Eqs. (15) and (22) are exactly the same. In short, for small angles the results of Table 1 not only reduce to those predicted by Einstein's geometrical theory, they are also  in agreement with the classical equivalence principle. 

\begin{figure}[t]
\centerline{\psfig{file=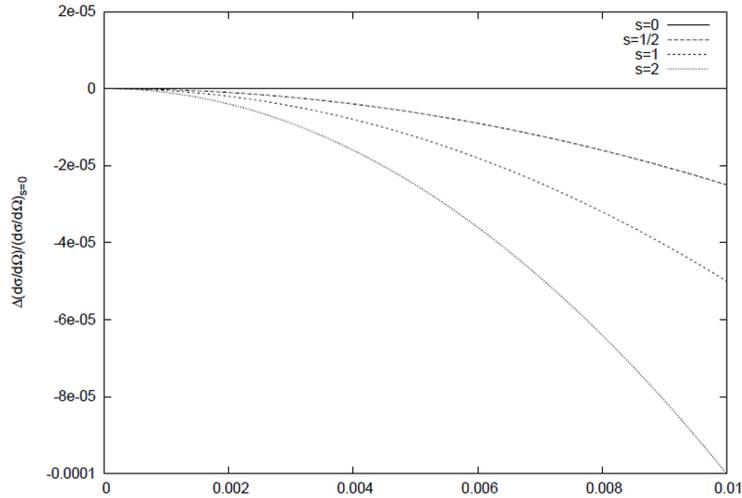,width=10cm}}
\caption{$\frac{\Delta(\frac{d\sigma}{d\Omega})}{(\frac{d\sigma}{d\Omega})_{s=0}}$ as a function of the scattering angle $\theta$.\label{fig2}}
\end{figure}}

\item{At first glance it seems that Eq. (16) predicts a dispersive deflection angle for the massive photons. Actually, this is a false impression; indeed, it is straightforward to show that Eq. (16) can be rewritten as 

\begin{eqnarray}            
\theta = \theta_\mathrm{E} \left( \frac{3 - {\bf v^2}}{2}\right).
\end{eqnarray}}
\end{itemlist}

After these important digressions, we return to the analysis of Eq. (16). The first term in the expression (16) coincides with that obtained by Einstein in 1916, by solving the equation of light propagation in the field of a static body,\cite{24} while the second one is the most important correction to the mass $m$ of the massive photon.

On the other hand, the angle of gravitational bending determined by the experimental groups is in general expressed through the relation\cite{25}

\begin{eqnarray}
\theta_\mathrm{exp} = \frac{1 + \gamma}{2} \theta_\mathrm{E},
\end{eqnarray}

\noindent where $\gamma$ is the deflection parameter, precisely and unambiguously determined by experimenters by measuring the deflection of electromagnetic radiation in the gravitational field of the sun.

From Eqs. (16) and (24), we then get

\begin{eqnarray}
m < 2\pi \nu \sqrt{\left|1- \gamma\right|}.
\end{eqnarray}

This amazing result clearly shows that there exists a constraint between the photon mass and the parameters $\gamma$ and $\nu$. In addition, it tells us that the knowledge of these parameters is all we need to set up upper limits on the photon mass.

\section{Finding Gravitational Upper Bounds on the Photon Mass}

We are now  finally ready  to find gravitational bounds on the photon mass. To accomplish this goal, we make use of the recent measurements of the solar gravitational deflection of radio waves found by Fomalont \emph{et al.}\cite{13} using the VLBI. The solutions for $\gamma$ obtained from these measurements,   as well as the corresponding bounds  on the photon mass we have estimated using Eq. (25),  are displayed   in Table 2. 

In order to determine the optimum value of $(\gamma -1)$ from the experimental results, the aforementioned authors minimized the (normalized) chi-squared expression

\begin{equation} 
\chi_k^2 = \frac{1}{k} \sum_{d,i}\left(\frac{P_d(i) - 0.5(\gamma -1)D_d(i)}{\sigma_d(i)}\right)^2,
\end{equation}

\noindent where $P_d(i)$ and $\sigma_d(i)$ are the measured position offset and error estimate, in this order. The term $D_d(i )$ is the differential general relativity gravitational bending prediction averaged over the session. The  $\sigma_\gamma$'s and $\chi_k^2$'s directly  related to the the solutions for $\gamma$ are shown in Table 2.  

\begin{table}[!h]
\tbl{Uppers bounds on the photon mass estimated using the solutions for $\gamma$ found by Fomalont \emph{et al.}$^{13}$ }
{\begin{tabular}{lcccc}
\\
\hline
\hline
\\
Solution Type &$(\gamma-1)\times10^{-4}$&$\sigma_{\gamma}\times10^{-4}$&$\chi^{2}_{k}$& $m \times10^{-11}$($MeV$)\\
\\
\hline
$43\;GHz$ data (corona-free) & -2.4 & 3.2 & 0.9 & 1.7 \\
\\
$43\;GHz$ data only & -1.0 & 2.6 & 2.2 & 1.1 \\
\\
$43\;GHz$ data only - Oct05& -3.2 & 2.8 & 1.1 & 2.0\\
\\
$23\;GHz$ data only - Oct05& -2.0 & 2.4 & 4.7 & 0.8\\
\hline
\end{tabular}}
\end{table}

In  Table 2, the $43\;GHz$ corona-free fit is the most accurate since it has the lowest $\chi^2$. This is due to   the lessening of some coronal effects, and the increase of the position errors. The two $43\;GHz$ only solutions (with no removal of the ionosphere reduction) show the effect of the Oct05 session that was  made relatively close to the sun; the first solution  was found using data obtained  during sessions that lasted several days, while the second one is based on data that were got within the space of one day only. Finally, the $23\;GHz$ only solution suggests that coronal refraction, which is four times larger than that at $43\;GHz$, is dominating the sensitivity of the experiment at $23\;GHz$.

 It is worth mentioning that Fomalont $et\;al.$\cite{13} have used for the results in their paper an average of the four solutions exhibited in Table 2 to obtain $\gamma=0.9998\pm0.0003$. From this result and assuming that the massive photon passing near the solar limb has a frequency $\nu=43 GHz$, which is perfectly justifiable since their data came mainly from $43 GHz$ observations where the refraction effects of the solar corona were negligible beyond  3 degrees from 
the sun, we obtain another gravitational bound on the photon mass, namely, $m \sim 3.4 \times 10^{-11} MeV$.

For reasons already discussed, we come to the conclusion that among the gravitational bounds on the photon mass we have found, the most reliable  is $m\sim1.7\times10^{-11}MeV$. In fact, the associated $\gamma$ was determined, on the one hand, using one single frequency; on the other, it has the best chi-squared. Furthermore, the data used for computing $\gamma$ came mainly from observations where the refraction effects of the solar corona were negligible.   

\section{Discussion}

Certainly, the bounds we have found  on the photon mass are considerably higher than the recently recommended limit published by the Particle Data Group ($m < 1 \times 10^{-18} eV$).\cite{26} They are nevertheless comparable to other existing bounds.$^{20-22}$ Let us then discuss whether or not  a better limit on the photon mass might be obtained using Eq. (25). First, if the deflections measured using the VLBI could be made with greater  accuracy the value of $\sqrt{\left|1 - \gamma\right|}$ would be reduced giving, as a result, a better  gravitational estimate. According to Fomalont {\it et al.},\cite{13} a series of designed experiments with the VLBI could increase the accuracy of the future experiments by at least a factor of 4. Second,  if deflection measurements can be obtained at lower frequencies, while maintaining  the value of the deflection parameter $\gamma$, the gravitational bound will be improved in direct proportion to  the frequency. This point, however, is very delicate. In fact, as we have already mentioned, up till now  the best results obtained for the gravitational deflection via the VLBI are those that  come mainly from $43 GHz$ where the refraction effects of the solar corona were negligible beyond 3 degrees from the sun. Incidentally, the lowest frequency employed by the radio astronomers was $2GHz$. However, the measurements made at this frequency are less reliable because of the refraction effects of the solar corona.   Actually, the radio astronomers use in their experiments a mixing of different frequencies but the most significant  contributions come in general from $\sim 43 GHz$. The possibility of improving  the gravitational limit on the photon mass in this case is thence very limited.

We remark that  up to now   only two attempts    to constrain   the photon mass using gravitational deflection measurements were made.\cite{27,28}   Unfortunately, these estimates are not very reliable because in both of them the values of the deflection parameter $\gamma$ are overestimated, while the values of the frequency are underestimated. In reality,  the $\nu$-values  used in the mentioned estimates are  in the neighborhood of the the lowest frequency employed by radio astronomers  ($\approx 2GHz$).

We call attention to the fact that the gravitational estimates on the photon mass we have obtained, like the great majority of estimates made with the purpose of limiting the photon mass and which are available in the literature, are essentially order-of-magnitude arguments. Nonetheless, our efforts in this work   are based on a new conceptual approach to the subject; besides, in the calculation  of the bounds, the most accurate experimental data currently available have been taken as inputs.

To conclude, we remark that recently we have found a quantum bound on the photon mass ($m \sim 1.6\times 10^{-10} MeV$) based on the computation of the anomalous electron magnetic moment in the framework of Proca electrodynamics.$^{29,30}$ To the best of our knowledge, this is the firs time a quantum bound on the photon mass is estimated.\footnote{Note that the bound on the photon mass obtained by Boulware and Deser$^{31}$ via the Aharonov-Bohm effect (which is present in massive (finite-range) electrodynamics) is a semiclassical limit; our bound, nonetheless, unlike the Bolware-Deser limit, is based on truly (loop) quantum effects.} It is worth noticing that this quantum bound is one order of magnitude higher than the bound derived from the gravitational scattering.

\section*{Acknowledgments}
The authors are very grateful to FAPERJ, CNPq, and CAPES (Brazilian funding agencies) for the financial support.

\end{document}